# Photothermoelectric Effects and Large Photovoltages in Plasmonic Au Nanowires with Nanogaps

Pavlo Zolotavin[1], Charlotte Evans[1], Douglas Natelson[*,1,2,3]

[1]Department of Physics and Astronomy, Rice University, 6100 Main St., Houston, Texas 77005, United States

[2]Department of Electrical and Computer Engineering, Rice University, 6100 Main St., Houston, Texas 77005, United States

[3]Department of Materials Science and NanoEngineering, Rice University, 6100 Main St., Houston, Texas 77005, United States

[*]E-mail: natelson@rice.edu.

**Abstract:** Nanostructured metals subject to local optical interrogation can generate open-circuit photovoltages potentially useful for energy conversion and photodetection. We report a study of the photovoltage as a function of illumination position in single metal Au nanowires and nanowires with nanogaps formed by electromigration. We use a laser scanning microscope to locally heat the metal nanostructures via excitation of a local plasmon resonance and direct absorption. In nanowires without nanogaps, where charge transport is diffusive, we observe voltage distributions consistent with thermoelectricity, with the local Seebeck coefficient depending on the width of the nanowire. In the nanowires with nanogaps, where charge transport is by tunneling, we observe large photovoltages up to tens of mV, with magnitude, polarization dependence, and spatial localization that follow the plasmon resonance in the nanogap. This is consistent with a model of photocurrent across the nanogap carried by the nonequilibrium, "hot" carriers generated upon the plasmon excitation.

**Table of Contents Figure:**

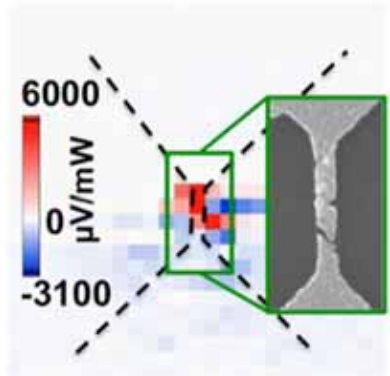

In the Seebeck effect, a diffusive conductor subject to a temperature gradient develops an open-circuit potential difference along its length, to counter the thermal diffusion of carriers.[1] When the temperature gradient is driven by optical excitation, this thermoelectric effect may be leveraged for photodetection.[2,3] Advances in the development of nanostructured materials have created new opportunities to improve thermoelectric response.[4,5] In thin metal films and nanowires, the proximity of the surfaces is detrimental to thermoelectricity, as surface scattering reduces electron and phonon mean free path, which lowers both electronic and "phonon drag" contributions to the Seebeck coefficient.[6–9] This sensitivity to boundary scattering makes it possible to create single-material thermocouples by controlling its geometry on the length scale of a few times the electronic mean free path.[10,11] The thermoelectric power of metals can also be modified by the creation of the atomic-scale junctions for which the Seebeck coefficient is determined by the energy dependence of the transmission properties of the junction.[12,13] Single-molecule junctions have been intensely studied recently following proposals for enhanced thermoelectric properties due to quantum interference effects,[14–18] though simple tunneling junctions are not expected to show particularly large Seebeck response.[19,20]

The conventional treatment of thermoelectric effects assumes that electron and lattice temperatures are the same; however this may not be the case in optically driven systems where light can generate short-lived, high energy carriers.[21–25] Hot electrons generated in plasmonic nanostructures were demonstrated to create additional photocurrent in molecular and tunnel junctions.[26–28] The effects of non-thermal electronic distributions on thermoelectric properties in plasmonic metal nanostructures have been predicted theoretically,[29] but only received experimental attention in the context of photothermoelectric (PTE) properties of graphene based photodetectors.[30,31] The connecting electrodes in photodetectors and other nanoscale devices are typically treated in a bulk-like manner without considering a possibility of the nanoscale variation of the Seebeck coefficient and the effects of non-thermal electronic distributions.[10,11,29,32]

With the advent of nanostructuring to modify thermoelectric response and plasmonic nanostructures to enhance hot electron effects, a systematic examination of locally heated metal nanostructures is key to understanding the physics behind the resulting photovoltages. Here we use a laser scanning microscope to heat metal nanostructures locally via direct absorption and, in nanogap devices, excitation of a highly localized multipolar plasmon modes present in the nanogap.[33]

We report a study of the photovoltage generated in thin gold nanowires at substrate temperatures between 5 K and room temperature as a function of illumination position. The voltage generated in short, unbroken nanowires is consistent with the PTE behavior observed in single metal thermocouples.[10,11] We also study the photoresponse of the nanowires containing a plasmonically active nanogap formed by electromigration. In these tunneling devices, we routinely observe photovoltages of tens of mV, which constitutes a large enhancement compared to the values expected in these nanostructures under the temperature gradient imposed by ordinary heating. We discuss several possible explanations for the observed behavior, emphasizing the role of hot electrons in the enhancement of the photovoltage in devices with tunneling nanogaps. Our results demonstrate unexplored opportunities for enhancing the thermoelectric properties of metal nanostructures and the performance of plasmonic photodetectors.

A typical unbroken bowtie device consists of the nanowire contacted by wide fan-out electrodes of the same metal, Figure 1a. The majority of the results reported here are for devices made of 14 nm thick gold with 1 nm titanium adhesion layer fabricated on thermally oxidized silicon wafers. For control experiments, we also studied devices fabricated without the Ti adhesion layer, devices using AuPd alloy or Ni instead of Au, and devices fabricated on sapphire substrates. As the laser beam is raster scanned across the bowtie device, it acts as a heat source and locally raises the temperature of the metal film, which creates an open circuit thermoelectric voltage measured by the lock-in amplifier, Figure 1b. The PTE voltage map is antisymmetric with respect to the center of the device, with maxima and minima located in the connecting fan-out electrodes close to the ends of the nanowire, Figure 1c. Qualitatively similar results are obtained for substrates at base temperature of 5 K, Figure 1d. The laser power dependence of the signal is linear at room temperature, Figure 1e, but slightly deviates from linearity at low temperature, Figure S2. Longitudinally polarized light (polarization angle is denoted as 0°) eliminates the contribution of the transverse plasmonic heating in the nanowire and leaves only the heating contribution from direct optical absorption. The resulting PTE voltage is similarly reduced, as seen in Figure 1f. A smaller magnitude of the PTE voltage is also observed in devices fabricated on sapphire substrate, Figure S3. The improved thermal dissipation through the substrate leads to a reduced optically driven temperature increase,[34] which further confirms the thermal origin of the induced voltages.

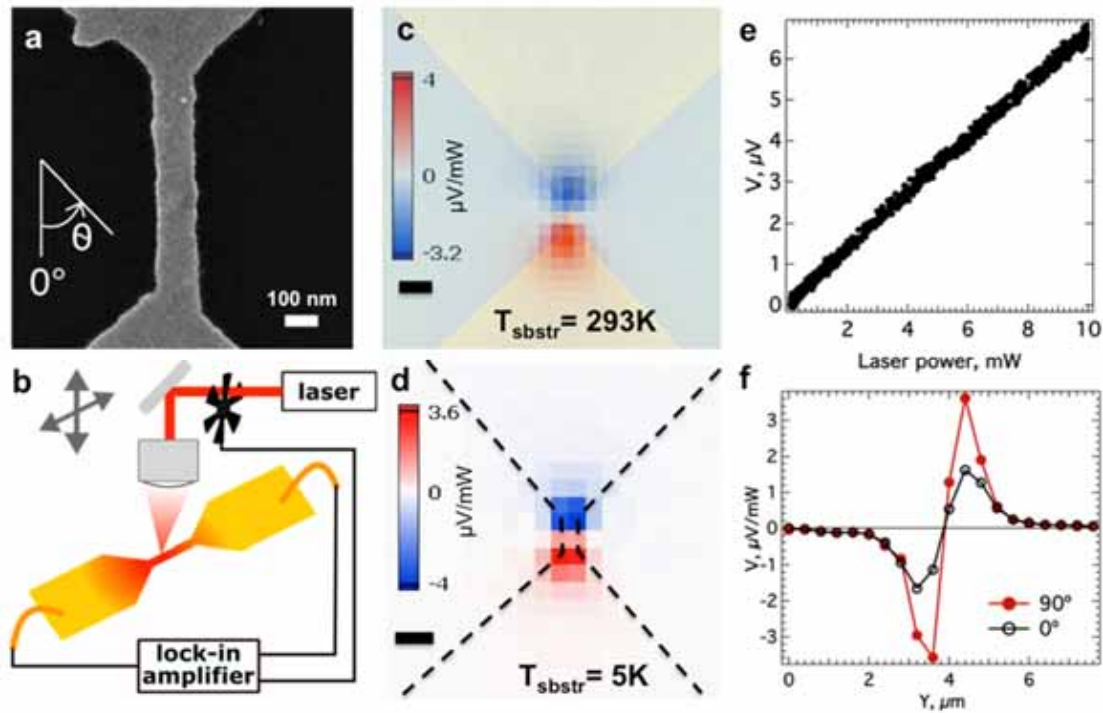

**Figure 1**. Photothermoelectric (PTE) voltage maps of short Au/Ti bowtie devices. (a) Scanning electron microscope (SEM) image of a typical bowtie device before electromigration. The width of the nanowire was kept close to 130 nm, which allows for the excitation of a transverse plasmon resonance with 785 nm laser polarized perpendicular to the long direction of the nanowire. (b) Schematics of the experimental setup with details provided in the Methods section. (c) Spatial map of photothermally generated voltage in Au/Ti nanowire at substrate temperature of 5

K. The map is superimposed with a false colored SEM image of the bowtie device. (d) PTE voltage map for a different device measured at substrate temperature of 5 K. Dashed line defines the bowtie contours. (e) Laser power dependence of the PTE voltage recorded in the location of the maximum of the PTE voltage map. Data acquired at room temperature. (f) Voltage along the vertical centerline of the device from (d) for the polarization perpendicular (filled circles, 90°) and parallel (empty circles, 0°) to the nanowire. For panels (c), (d), and (f) voltage is reported in units of µV per mW of laser power on the sample. The bottom electrode is connected to the -B input of the voltage amplifier. Scale bar in (c), (d) is 1 µm.

Using a bolometric approach and computational modeling, we have previously inferred the temperature rise of the nanowires due to the plasmon resonance excitation in similar experimental conditions.[34,35] The voltages generated at room temperature, Figure 1c, and at 5 K, Figure 1d, are comparable in magnitude. This coincidental result stems from the fact that PTE voltage is proportional to the product of the Seebeck coefficient and local temperature increase. At room temperature, the small temperature increase, $\Delta T \sim 10$ K, is multiplied by a relatively large room temperature value of the Seebeck coefficient for gold, $S_{Au} \sim 1.5$ µV/K. At low temperature, the decreased $S_{Au}$ is compensated by a significantly larger $\Delta T \sim 140$ K at similar laser power level. The magnitude and spatial distribution of the PTE voltages in short nanowires are consistent with the spatial variability of Seebeck coefficient determined by the width of the device; see Supplementary Information (SI) for details of the estimate calculated using COMSOL Multiphysics.

The amplitude and sign of the closed circuit photocurrent generated due the PTE voltage present in the device is consistent with the voltage maps acquired in a pair of sequential scans Figure 2a,d. Deliberate formation of a nanoscale constriction in the nanowire by the onset of the electromigration leads to a mild increase in the magnitude of the PTE voltage, Figure 2b,e. In this particular device, the constriction is formed closer to the bottom electrode, breaking the symmetry for thermal dissipation in the nanostructure, Figure S4. The local temperature increase in the nanowire in now larger and is not symmetric around the break point due to the reduced thermal conductance through the constriction, which leads to a change in the PTE voltage map, Figure 2b,e. When the device resistance is increased beyond that of quantum of resistance, $h/2e^2 \approx 12.9$ k$\Omega$, the nanogap is formed and the magnitude of the photovoltage is substantially increased, Figure 2c. The closed circuit photocurrent continues to mirror the photovoltage map for the devices with a nanogap, Figure 2f.

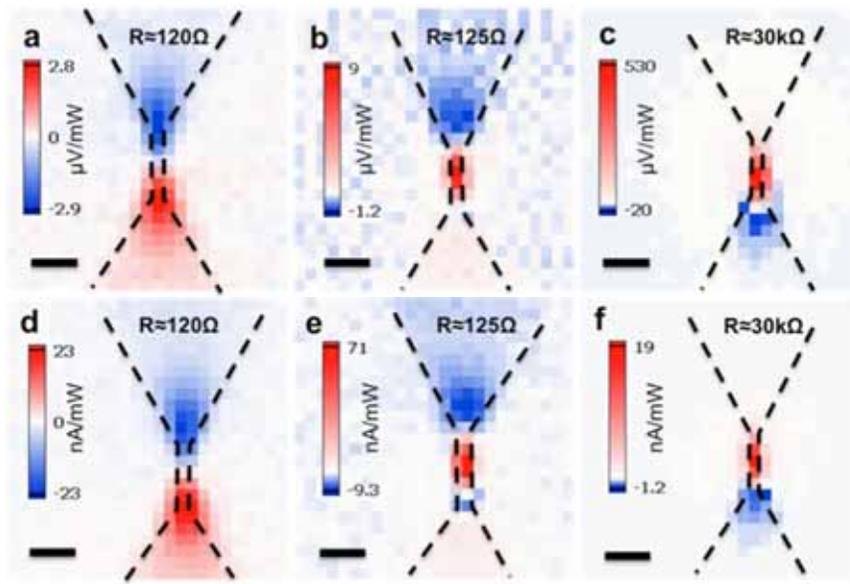

**Figure 2**. PTE voltage maps (top row) and the corresponding PTE current maps (bottom row) for the Au/Ti device recorded during the formation of the nanogap by electromigration. Data were acquired at substrate temperature of 293K and 1mW incident laser power. The scale bar is 1 μm.

Electromigration of the nanowires at low temperatures improves device stability at intermediate resistance values. The trend of modest increase in the magnitude of the PTE voltage after the constriction formation was observed in all devices studied and usually saturates by the time device resistance reaches ~1 k$\Omega$. Evolution of the PTE voltage map, for another device as it is electromigrated at substrate temperature of 5 K, is shown in Figure 3. Before the formation of the nanogap, the PTE voltage of this wire remains small and the spatial distribution of the PTE voltage does not change as device resistance is increased, as the constriction forms closer to the center of this particular nanowire.

After the tunneling nanogap is formed, the data make clear that a different physical mechanism is at work, rather than diffusive Seebeck response. The photovoltage is further increased by ~20×, Figure 3d. The positions of voltage minimum and maximum are now located closer to the nanogap rather then at the ends of the nanowire, without a resolvable zero voltage region between them. The voltage maps can vary from scan to scan not only by spatial distribution of voltage sign, but also by magnitude, Figure 3e, demonstrating additional ~30× increase in observed signal for this particular device. This temporal variation, worse at high substrate temperatures and high incident optical powers, shows that the mechanism responsible for photovoltages in tunneling nanogaps depends on atomic-scale details of the gap region. In total, with a nanogap we observe a ~1000× photovoltage enhancement compared to the initial PTE voltage in the diffusive devices with no nanogap. The enhanced open circuit photovoltage remains present after the device is warmed to room temperature, Figure 3f. The device-to-device variation is large after the nanogap formation, but photovoltages on the order of tens of mV are routinely observed both at low and room temperature experiments regardless of the presence of the Ti adhesion layer, Figure S5, S6. Figure 3 shows performance of a typical device.

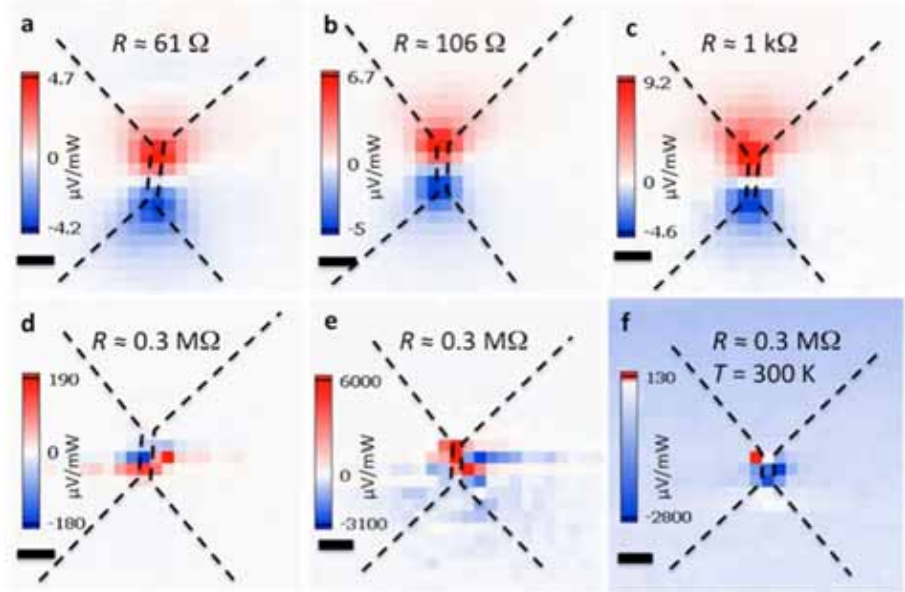

**Figure 3**. Evolution of the photovoltage map as a Ti-free Au device is electromigrated to form a nanogap. Panels (a)-(e) are at a substrate temperature of 5 K. (a) Initial device resistance is 61 Ω. (b) The same device after some electromigration to a higher resistance constriction. Positive PTE voltage is increased ~2×, but the negative side remains approximately of the same magnitude. (c) The junction is further migrated to 1 kΩ. This asymmetry in the PTE voltage map remains and is the consequence of the electromigration creating the constriction closer to the grounded fan-out electrode (in this case lower side of the junction). (d) The junction after nanogap formation with a measured resistance of 0.3 MΩ. (e) Further enhancement of the photovoltage after several scans. (f) The same device after warm up to room temperature at laser power of 1 mW. Units for all panels are in μV per mW of laser power on the sample. Scale bars are 1 μm.

The "hot spot" in the photovoltage maps after nanogap formation is extremely localized. The sign of the voltage can vary between adjacent pixels even though the spatial separation between the pixels is only 0.3 μm, which is much smaller than the laser spot size. As mentioned, the exact spatial distribution of photovoltages is typically not reproducible between consecutive laser scans at high laser power (10 mW), but is more stable at lower laser power levels (1 mW or lower), Figure 4a. The increased device stability allows recording polarization dependence of the photovoltage, which displays a dipole character with maximum response at a polarization that coincides with the transverse plasmon excitation, Figure 4b.

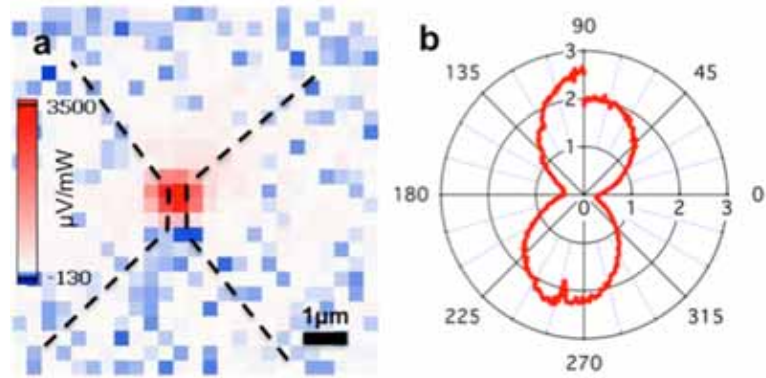

**Figure 4**. Improved stability and polarization dependence of enhanced photovoltage at low laser power for the device demonstrated in Figure 3. (a) Photovoltage map at 1 mW incident laser power recorded at substrate temperature of 5 K. (b) Polarization dependence of the photovoltage recorded in the maximum of (a). Units are in mV per mW of laser power on the sample for (b). Polarization dependence does not perfectly retrace to the initial value at the end of the scan due to temporal drift in sample position during the measurement.

The PTE voltage in devices *without* the nanogap is produced by the conventional thermoelectric effect in metal films. Localized heating creates an area of the nanostructure with a slightly higher steady state electronic temperature. A potential difference then develops across the device to counter the migration of the high-energy carriers to the cold section because there cannot be net electrical current in the open circuit configuration. The shape of the Fermi energy surface and details of the electron diffusion through metal film determine the magnitude of the thermoelectric effect in this case.

For the plasmonic nanowires *with* the nanogap, we argue that the origin of the photovoltage is also caused by the difference in the electron energy distribution in the two sides of the device, and the development of an open-circuit potential to null out net current. Transport in this case, however, is through tunneling, and it is the deviation of the distributions from the thermal form due to the photoexcitation of high-energy carriers that is key. Following the model developed in Ref. 36, we use the Landauer theory expression to define the current across the nanogap

$$I_p = \frac{2e}{h} \int_{-\infty}^{\infty} d\varepsilon \, [f_1(\varepsilon) - f_2(\varepsilon)] \mathcal{T}(\varepsilon),$$

where $e$ is electron charge, $f_j(\varepsilon)$ ($j = 1,2$) is the electron distribution in electrode $j$ and $\mathcal{T}(\varepsilon)$ is the transmission function at electron energy $\varepsilon$. Assuming that the electrode 1 is predominantly illuminated, we can write $f_1(\varepsilon, t) - f_2(\varepsilon) = f_{eq}^{(1)}(\varepsilon, t) - f_{eq}^{(2)}(\varepsilon) + f_p^{(1)}(\varepsilon, t)$, where the difference $f_{eq}^{(1)}(\varepsilon, t) - f_{eq}^{(2)}(\varepsilon)$ corresponds to the difference in equilibrium electronic temperatures of electrodes and accounts for conventional thermoelectric effects, $f_p^{(1)}(\varepsilon, t)$ is the non-equilibrium component induced by light absorption. In our case of CW illumination, the photo-induced current could be written as

$$I_p = \frac{2e}{h} \frac{\dot{N}_I}{hf} L(f_I) \int_{-\infty}^{\infty} d\varepsilon \frac{\tau_e(\varepsilon)}{\rho(\varepsilon)} \frac{\tanh\left(\frac{\varepsilon \beta_\varepsilon}{2}\right)}{1+e^{-\beta_\varepsilon h f_I} \cosh(\varepsilon \beta_\varepsilon)} \mathcal{T}(\varepsilon),$$

where $\dot{N}_I$ is the number of electrons per second per atom, $L(f_I)$ is the absorption lineshape at incident frequency, $\tau_e(\varepsilon)$ excess distribution relaxation time, $\rho(\varepsilon)$ is the density of states per atom, and $\beta_e = (k_B T)^{-1}$. In the open circuit configuration, the voltage $V_p$ needed to nullify the "hot" electron current component is then set by

$$I_p = -\left(\frac{2e^2}{h}\right) \mathcal{T}(\varepsilon = 0) V_p$$

Beside material specific parameters, the details of the transmission through the nanogap will determine the magnitude of the $I_p$ and ultimately the observed photovoltage, $V_p$.

The tunneling in the nanogaps is likely to be mediated by adsorbates of unknown origin present in the gap, Fig. S9. The simplest model of the junction with transmission dominated by a single conduction resonance $\mathcal{T}(\varepsilon) = \Gamma^2/[(\varepsilon-\varepsilon_0)^2+\Gamma^2]$, where $\varepsilon_0$ is the resonance energy and $\Gamma$ is the electrode coupling, allows $I_p$ to be estimated provided $\varepsilon_0$ and $\Gamma$ are known. The latter could be estimated from the nanogap resistance without laser illumination. As a typical example, we consider a 10MΩ device and typical values for electrode coupling $\Gamma \sim 0.1$ we estimate the resonance energy $\varepsilon_0 \sim 2.7 eV$. For a 785nm CW laser with intensity of 400kW/cm² (which corresponds to 10mW of laser power) $\dot{N}_I \sim 1.6*10^9$ photons per sec per atom. Using a free electron model, the density of states per atom, $\rho(\varepsilon)$, is $\sim 0.1$eV$^{-1}$. From a COMSOL calculation of the scattering cross section of the nanowire, we deduce $L(f_I) \sim 0.05$. Taking a conservative estimate of $\tau_e(\varepsilon) \sim 10^{-14}s$, we estimate $I_p \sim 0.5 nA$, which would correspond to the photovoltage of 5mV – a value typically observed experimentally. Similar or larger values of the "hot" electron photocurrent in molecular junctions with plasmonic electrodes were obtained by other groups.[26–28]

The hot-carrier current is exponentially dependent on the position of the resonance energy level and the effective equilibrium electron temperature, which can explain device-to-device variability of the PTE voltage. Excitation of the localized plasmon resonance in metal nanostructures can create populations of relatively long-lived high-energy carriers that, while not thermally distributed, can produce effects similar to an increased local electronic temperature.[37,38] The intense spatial variability of the nanogap photovoltage signal is consistent with the local nature of the plasmon enhancement in the electromigrated nanogaps.[33] Formation of dark plasmon modes that lead to the field enhancement in these structures depends on the minute structural details of the surface of the nanogap, and excitation efficiency of those modes can depend sensitively on spot position. Moreover, the exact local details of the geometry can change under laser illumination, leading to changes in the local plasmonic properties of the nanogap creating temporal variability of the photovoltage in the "hot" spot. Strong polarization dependence is also consistent with a plasmonic origin of the observed photovoltage. The control experiments in AuPd and Ni devices, Figure S7, S8, S10 that do not possess resonant localized plasmonic modes display nanogap photovoltages that are considerably weaker, though still present. Even in the absence of plasmon dissipation, light absorption can create short-lived high-energy carriers that can transit the nanogap and lead to the development of a photovoltage, albeit to a smaller degree.

A number of other physical processes could contribute to the large photovoltages in the nanogaps: optical rectification,[25,39,40] photo-assisted electron tunneling,[41] the plasmoelectric effect,[42] and thermoelectric effects in adsorbates or contaminants in the nanogap, Fig S8. In the SI we discuss each of these alternatives and the reasons why we consider them of lesser importance or irrelevant. The optical creation of the high-energy carriers and their nonequilibrated transport is a mechanism similar to that seen in graphene-based photodetectors, and plausibly consistent with the observed systematics of the enhanced PTE voltages in plasmonic nanogaps.

In conclusion, we study photovoltage properties of illuminated plasmonic bowtie devices. In the nanowires without a nanogap, photovoltages are small and consistent with photothermoelectricity due to local optical heating and the spatially dependent Seebeck coefficient. In nanowires with nanogaps, we observe greatly enhanced photovoltages under resonant laser illumination. While detailed modeling is very challenging, observations are consistent with estimates based on a model in which the potential arises to null tunneling of photogenerated hot carriers. The magnitude of the large nanogap photovoltage response and the role of plasmons suggested by the systematics of the experimental results imply that there are new opportunities for plasmonic tunneling devices as photodetectors. Deliberate engineering of structures to favor plasmonic hot electron production in proximity to tunnel junctions, and the integration of materials with larger Seebeck coefficients in such devices, would test these ideas and be new tools in the study of hot carrier generation. Control of the geometry of the device at the nanoscale might offer new tools for controlling both thermoelectric and hot electron response.

**Methods**

The bowtie devices were fabricated using standard e-beam lithography techniques on Si wafers with a 200 nm of thermally grown oxide layer. For sapphire substrates a 6 nm chromium layer was evaporated on top of the polymer resist to reduce charging effects during e-beam writing. Larger Au/Ti contact pads for wire bonding were shadow mask evaporated prior to the device fabrication. A sample containing 24 devices with shared ground was wire bonded to a larger chip carrier. The results demonstrated here were selected to represent the typical behavior. Experiments were carried out in a home-built Raman microscope set up.[34] The total number of devices examined is 185. The devices were kept under high vacuum in the closed-cycle optical cryostat (Montana Instruments). Laser light was modulated using a mechanical chopper at 287 Hz. The PTE voltage maps are acquired with laser polarization perpendicular to the long dimensions of the nanowire (90°) unless specially noted. Thermal voltage was measured using SR560 voltage amplifier connected to a lock-in amplifier synced to a chopper frequency. For devices with a nanogap formed by electromigration, the voltage was measured directly using NI DAQ without laser light modulation. Unless specified, measurements were performed at laser power of 10mW measured at the sample. Maps of the closed circuit photocurrent were measured using SR570 current amplifier.

**Supporting Information Available:** 2D COMSOL heat transfer model, additional experimental data, and discussion of other physical mechanisms that can create photovoltages in nanogaps. The material is available free of charge via the Internet at http://pubs.acs.org.

**Acknowledgments** P.Z. and D.N. acknowledge support from ARO award W911-NF-13-0476 and from the Robert A. Welch Foundation Grant C-1636. C.E. acknowledges support from NSF GRFP DGE-1450681.

**Competing financial interests**. The authors declare no competing financial interests.

# References

(1) Blatt, F. J.; Schroeder, P. A.; Foiles, C. L.; Greig, D. *Thermoelectric Power of Metals*; Springer US: Boston, MA; 1976.
(2) Cai, X.; Suess, R. J.; Drew, H. D.; Murphy, T. E.; Yan, J.; Fuhrer, M. S. Pulsed Near-IR Photoresponse in a Bi-Metal Contacted Graphene Photodetector. *Sci. Rep.* **2015**, *5,* 14803.
(3) Szakmany, G. P.; Orlov, A. O.; Bernstein, G. H.; Porod, W. Shape Engineering of Antenna-Coupled Single-Metal Nanothermocouples. *Infrared Phys. Technol.* **2015**, *72*, 101–105.
(4) Snyder, G. J.; Toberer, E. S. Complex Thermoelectric Materials. *Nat. Mater.* **2008**, *7*, 105–114.
(5) Vineis, C. J.; Shakouri, A.; Majumdar, A.; Kanatzidis, M. G. Nanostructured Thermoelectrics: Big Efficiency Gains from Small Features. *Adv. Mater.* **2010**, *22*, 3970–3980.
(6) Lin, S. F.; Leonard, W. F. Thermoelectric Power of Thin Gold Films. *J. Appl. Phys.* **1971**, *42*, 3634–3639.
(7) Cattani, M.; Salvadori, M. C.; Vaz, A. R.; Teixeira, F. S.; Brown, I. G. Thermoelectric Power in Very Thin Film Thermocouples: Quantum Size Effects. *J. Appl. Phys.* **2006**, *100*, 114905.
(8) Shapira, E.; Tsukernik, A.; Selzer, Y. Thermopower Measurements on Individual 30 Nm Nickel Nanowires. *Nanotechnology* **2007**, *18*, 485703.
(9) Avery, A. D.; Sultan, R.; Bassett, D.; Wei, D.; Zink, B. L. Thermopower and Resistivity in Ferromagnetic Thin Films near Room Temperature. *Phys. Rev. B* **2011**, *83*, 100401.
(10) Sun, W.; Liu, H.; Gong, W.; Peng, L.-M.; Xu, S.-Y. Unexpected Size Effect in the Thermopower of Thin-Film Stripes. *J. Appl. Phys.* **2011**, *110*, 83709.
(11) Szakmany, G. P.; Orlov, A. O.; Bernstein, G. H.; Porod, W. Single-Metal Nanoscale Thermocouples. *IEEE Trans. Nanotechnol.* **2014**, *13*, 1234–1239.
(12) Ludoph, B.; Ruitenbeek, J. M. van. Thermopower of Atomic-Size Metallic Contacts. *Phys. Rev. B* **1999**, *59*, 12290–12293.
(13) Evangeli, C.; Matt, M.; Rincón-García, L.; Pauly, F.; Nielaba, P.; Rubio-Bollinger, G.; Cuevas, J. C.; Agraït, N. Quantum Thermopower of Metallic Atomic-Size Contacts at Room Temperature. *Nano Lett.* **2015**, *15*, 1006–1011.
(14) Reddy, P.; Jang, S.-Y.; Segalman, R. A.; Majumdar, A. Thermoelectricity in Molecular Junctions. *Science* **2007**, *315*, 1568–1571.
(15) Widawsky, J. R.; Darancet, P.; Neaton, J. B.; Venkataraman, L. Simultaneous Determination of Conductance and Thermopower of Single Molecule Junctions. *Nano Lett.* **2012**, *12*, 354–358.
(16) Rincón-García, L.; Evangeli, C.; Rubio-Bollinger, G.; Agraït, N. Thermopower Measurements in Molecular Junctions. *Chem. Soc. Rev.* **2016**, *45*, 4285–4306.
(17) Dubi, Y.; Di Ventra, M. Heat Flow and Thermoelectricity in Atomic and Molecular Junctions. *Rev. Mod. Phys.* **2011**, *83*, 131–155.
(18) Bergfield, J. P.; Solis, M. A.; Stafford, C. A. Giant Thermoelectric Effect from Transmission Supernodes. *ACS Nano* **2010**, *4*, 5314–5320.

(19) Leavens, C. R.; Aers, G. C. Vacuum Tunnelling Thermopower: Normal Metal Electrodes. *Solid State Commun.* **1987**, *61*, 289–295.
(20) Marschall, J.; Majumdar, A. Charge and Energy Transport by Tunneling Thermoelectric Effect. *J. Appl. Phys.* **1993**, *74*, 4000–4005.
(21) Koppens, F. H. L.; Mueller, T.; Avouris, P.; Ferrari, A. C.; Vitiello, M. S.; Polini, M. Photodetectors Based on Graphene, Other Two-Dimensional Materials and Hybrid Systems. *Nat. Nanotechnol.* **2014**, *9*, 780–793.
(22) Buscema, M.; Barkelid, M.; Zwiller, V.; van der Zant, H. S. J.; Steele, G. A.; Castellanos-Gomez, A. Large and Tunable Photothermoelectric Effect in Single-Layer MoS2. *Nano Lett.* **2013**, *13*, 358–363.
(23) Kallatt, S.; Umesh, G.; Bhat, N.; Majumdar, K. Photoresponse of Atomically Thin MoS2 Layers and Their Planar Heterojunctions. *Nanoscale* **2016**, *8*, 15213–15222.
(24) Sierra, J. F.; Neumann, I.; Costache, M. V.; Valenzuela, S. O. Hot-Carrier Seebeck Effect: Diffusion and Remote Detection of Hot Carriers in Graphene. *Nano Lett.* **2015**, *15*, 4000–4005.
(25) Shi, S.-F.; Xu, X.; Ralph, D. C.; McEuen, P. L. Plasmon Resonance in Individual Nanogap Electrodes Studied Using Graphene Nanoconstrictions as Photodetectors. *Nano Lett.* **2011**, *11*, 1814–1818.
(26) Fung, E.-D.; Adak, O.; Lovat, G.; Scarabelli, D.; Venkataraman, L. Too Hot for Photon-Assisted Transport: Hot-Electrons Dominate Conductance Enhancement in Illuminated Single-Molecule Junctions. *Nano Lett.* **2017**, *17*, 1255–1261.
(27) Vadai, M.; Selzer, Y. Plasmon-Induced Hot Carriers Transport in Metallic Ballistic Junctions. *J. Phys. Chem. C* **2016**, *120*, 21063–21068.
(28) Pal, P. P.; Jiang, N.; Sonntag, M. D.; Chiang, N.; Foley, E. T.; Hersam, M. C.; Van Duyne, R. P.; Seideman, T. Plasmon-Mediated Electron Transport in Tip-Enhanced Raman Spectroscopic Junctions. *J. Phys. Chem. Lett.* **2015**, *6*, 4210–4218.
(29) Yang, R.; Narayanaswamy, A.; Chen, G. Surface-Plasmon Coupled Nonequilibrium Thermoelectric Refrigerators and Power Generators. *J. Comput. Theor. Nanosci.* **2005**, *2*, 75–87.
(30) Gabor, N. M.; Song, J. C. W.; Ma, Q.; Nair, N. L.; Taychatanapat, T.; Watanabe, K.; Taniguchi, T.; Levitov, L. S.; Jarillo-Herrero, P. Hot Carrier–Assisted Intrinsic Photoresponse in Graphene. *Science* **2011**, *334*, 648–652.
(31) Wu, D.; Yan, K.; Zhou, Y.; Wang, H.; Lin, L.; Peng, H.; Liu, Z. Plasmon-Enhanced Photothermoelectric Conversion in Chemical Vapor Deposited Graphene P–n Junctions. *J. Am. Chem. Soc.* **2013**, *135*, 10926–10929.
(32) Lee, E.-S.; Cho, S.; Lyeo, H.-K.; Kim, Y.-H. Seebeck Effect at the Atomic Scale. *Phys. Rev. Lett.* **2014**, *112*, 136601.
(33) Herzog, J. B.; Knight, M. W.; Li, Y.; Evans, K. M.; Halas, N. J.; Natelson, D. Dark Plasmons in Hot Spot Generation and Polarization in Interelectrode Nanoscale Junctions. *Nano Lett.* **2013**, *13*, 1359–1364.
(34) Zolotavin, P.; Alabastri, A.; Nordlander, P.; Natelson, D. Plasmonic Heating in Au Nanowires at Low Temperatures: The Role of Thermal Boundary Resistance. *ACS Nano* **2016**, *10*, 6972–6979.

(35) Herzog, J. B.; Knight, M. W.; Natelson, D. Thermoplasmonics: Quantifying Plasmonic Heating in Single Nanowires. *Nano Lett.* **2014**, *14*, 499–503.
(36) Kornbluth, M.; Nitzan, A.; Seideman, T. Light-Induced Electronic Non-Equilibrium in Plasmonic Particles. *J. Chem. Phys.* **2013**, *138*, 174707.
(37) Brongersma, M. L.; Halas, N. J.; Nordlander, P. Plasmon-Induced Hot Carrier Science and Technology. *Nat. Nanotechnol.* **2015**, *10*, 25–34.
(38) Knight, M. W.; Sobhani, H.; Nordlander, P.; Halas, N. J. Photodetection with Active Optical Antennas. *Science* **2011**, *332*, 702–704.
(39) Ward, D. R.; Hüser, F.; Pauly, F.; Cuevas, J. C.; Natelson, D. Optical Rectification and Field Enhancement in a Plasmonic Nanogap. *Nat. Nanotechnol.* **2010**, *5*, 732–736.
(40) Arielly, R.; Ofarim, A.; Noy, G.; Selzer, Y. Accurate Determination of Plasmonic Fields in Molecular Junctions by Current Rectification at Optical Frequencies. *Nano Lett.* **2011**, *11*, 2968–2972.
(41) Ittah, N.; Noy, G.; Yutsis, I.; Selzer, Y. Measurement of Electronic Transport through 1G0 Gold Contacts under Laser Irradiation. *Nano Lett.* **2009**, *9*, 1615–1620.
(42) Sheldon, M. T.; Groep, J. van de; Brown, A. M.; Polman, A.; Atwater, H. A. Plasmoelectric Potentials in Metal Nanostructures. *Science* **2014**, *346*, 828–831.

*Supporting Information*

# Photothermoelectric Effects and Large Photovoltages in Plasmonic Au Nanowires with Nanogaps

Pavlo Zolotavin[1], Charlotte Evans[1], Douglas Natelson[*,1,2,3]

[1]Department of Physics and Astronomy, Rice University, 6100 Main St., Houston, Texas 77005, United States

[2]Department of Electrical and Computer Engineering, Rice University, 6100 Main St., Houston, Texas 77005, United States

[3]Department of Materials Science and NanoEngineering, Rice University, 6100 Main St., Houston, Texas 77005, United States

[*]E-mail: natelson@rice.edu.

# 1. Estimate of the total voltage using a simplified 2D model of heating in the bowtie nanostructures

To generate nonzero thermoelectric voltage in a bulk wire made from a single material that is heated somewhere in the middle, while the ends are kept at identical temperatures, requires special circumstances.[1,2] The voltage difference between the ends of the wire (at $x = \pm l$) in this scenario is $V = \int_{-l}^{l} S(x,T)\nabla T(x)dx$, where $S$ is the location- and temperature-dependent Seebeck coefficient of the material and $T$ is the local temperature. On the nanoscale, $S$ can be locally modified by the change of the density of states as in atomic scale junctions[3,4] or electron mean free path as in single metal thermocouples.[5–9] In the latter case, the Seebeck coefficient of thin metal films is width-dependent. In bowtie devices, Figure 1a, when the laser beam is positioned in the middle of the nanowire, both the temperature profile and the geometry of the nanostructure are symmetric and therefore no PTE voltage is observed (Figure 1c,d). Where the nanowire transitions to the fan-out electrode, the increasing width of the device creates a small difference between the Seebeck coefficients of the nanowire and the fan-out electrode, which generates a PTE voltage when the beam is positioned in this spot. The sign of the PTE voltage depends on which side of the device is heated.

A spatial distribution of the PTE voltage across the device could be reproduced using a simplified 2D heat dissipation model that was implemented using a COMSOL Multiphysics. Heat dissipation was modeled using a 2D geometry with an out-of-plane heat transfer to the $SiO_2$ substrate. The left and right boundaries were kept at fixed temperature, 292 K, and the rest were set as insulating. The localized heat source had a Gaussian distribution to simulate heating from the focused laser beam. Additional heating from the plasmon resonance excitation was added as the width dependent heater modulation with the maximum contribution in the nanowire geometry segment. The intensity of the heat source was adjusted to produce a local temperature increase comparable to that measured experimentally, $\Delta T \sim 10$ K, Figure S1b. Simultaneously with the heat dissipation, the electric potential distribution was calculated. The right boundary of the device was grounded and the rest electrically isolated. The position of the heat source was moved along the centerline of the device to reproduce scanning of the laser beam and the voltage at the left boundary was recorded, Figure S1c. Using the Fuchs-Sondheimer electronic specular reflection model of the resistivity in thin films,[10,11] modified to accommodate reflection from the side walls, the Seebeck coefficient of the film could be written as[12–16]

$$S_f = S_g \left[1 - \frac{3}{8}\frac{(1-p)\lambda_0}{d}\frac{U_g}{1+U_g}\right]$$

where $d$ is the effective film thickness defined as $\frac{1}{d} = \frac{1}{w} + \frac{1}{t}$, $w$ is the width of the nanostructure and $t$ is the thickness of the film, $\lambda_0$ is the electron mean free path in gold, $p$ is the scattering coefficient, $U_g = \left.\frac{\partial \ln \lambda_0}{\partial \ln E}\right|_{E_F}$ and $S_g(T)$ is the Seebeck coefficient of the infinitely thick film approximated here by the bulk value.[17] We estimate the scattering coefficient $p$ as 0.1 by comparison of the resistivity and temperature coefficient of resistance with the previous data for thin films.[15] The value of $U_g$ for thin gold films is close to -0.6.[15] The above equation for $S_f$ is derived in the limit of $d \gg \lambda_0$, however in our case $d \sim 0.4\lambda_0$ and the prefactor 3/8 should therefore be reduced to 0.22. In this model the spatial dependence of the Seebeck coefficient is determined by the dependence of the electron mean free path on the width of the nanostructure. The model produces a spatial distribution of thermoelectric voltage that is qualitatively consistent with the one observed experimentally, Figure 1b of the main text. The location of the absolute maximum (minimum) of the thermoelectric voltage is determined by the position of the largest $\Delta S$ (for our $S(w,T)$ model it is located close to the nanowire ends as $dS \sim \frac{dw}{w^2}$) and the location of the $\Delta T$ maximum inferred from the heat dissipation calculation. A notable difference with the experimental data is that in the calculation the voltage minimum is located farther from to the nanowire end. This discrepancy is most likely due to the simplified nature of the model.

When comparing PTE maps for unbroken devices fabricated from Au, AuPd alloy, and Ni we see that magnitude of the PTE voltage generally follows the Seebeck coefficient of the metal used in the experiment. Close to room temperature the 60/40 AuPd alloy[18] has the largest Seebeck coefficient, $S_{AuPd}$ ~ -35 µV/K, and the largest signal ~ 100 µV/mW, Figure S7e. This is followed by Ni devices with $S_{Ni}$ ~ -20 µV/K and intermediate PTE voltage of ~ 50 µV/mW, Figure S8a. Au devices without Ti adhesion layer have the smallest PTE voltage of ~2-5 µV/mW and the smallest $S_{Au}$ ~ -1.5 µV/K. We note that AuPd devices fabricated with Ti adhesion layer had anomalously small value of the PTE voltage, which is most likely the result of the alloying of AuPd with Ti metal.

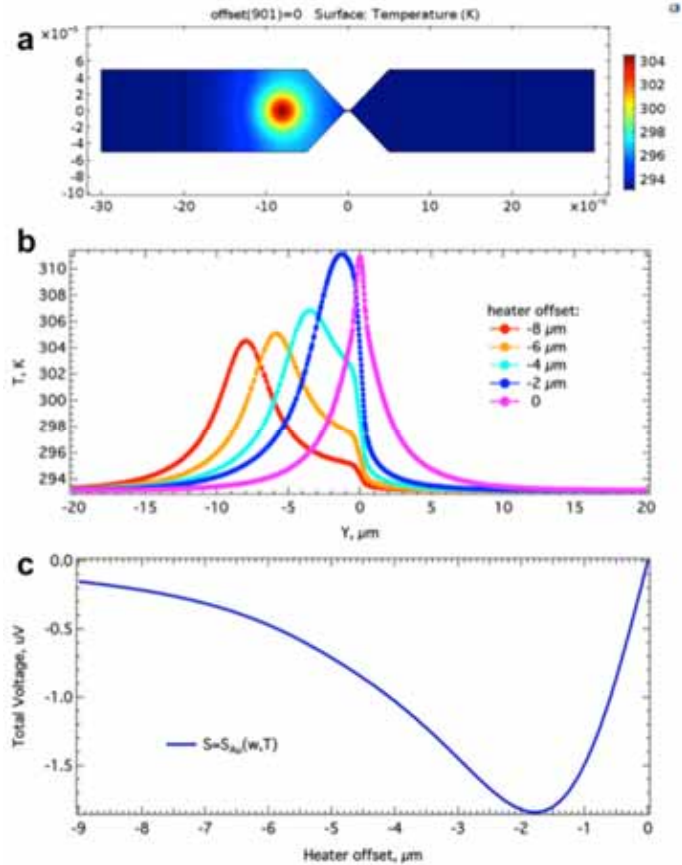

**Figure S1** Qualitative estimate of the total voltage measured in the experiment using a simplified 2D heating model with an artificial heater. (a) Surface temperature map of the temperature increase for the heater offset by -8 µm from the center of the device. The heater power was adjusted to produce ΔT ~ 10 K. The heater has a modified Gaussian spatial distribution to imitate localized heating from the focused laser beam and plasmon excitation. (b) Temperature profiles along the centerline of the device for different heater offsets. (c) Thermally generated voltage across the device as the function of the heater offset from the center of the device calculated using a spatially dependent Seebeck coefficient.

## 2. Additional experimental data for devices with unbroken nanowires

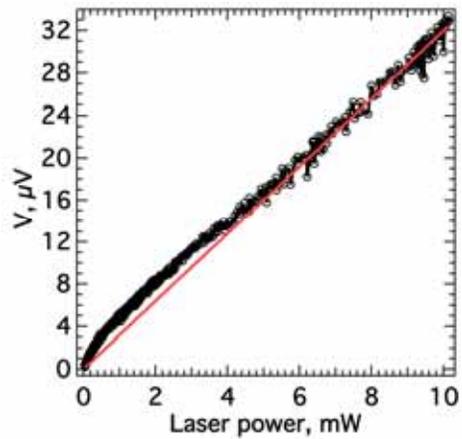

**Figure S2**. Laser power dependence of PTE voltage deviates from linearity at low temperatures. The non-linearity is a result of the combination of the non-linear laser intensity dependence of temperature increase of the nanostructure at low temperatures[19] and non-linear temperature dependence of Seebeck coefficient in the 5 K to 150 K. Data acquired at substrate temperature of 5 K.

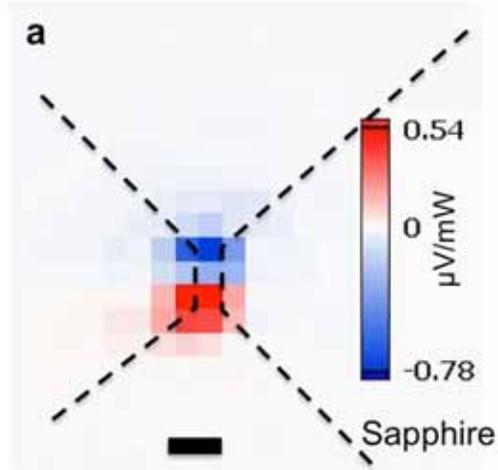

**Figure S3**. PTE voltage maps for Au/Ti devices fabricated on sapphire substrates. Data acquired at substrate temperature of 5 K and laser power of 10 mW on the sample. These experimental parameters correspond to $\Delta T \sim 45$ K in the center of the nanowire.

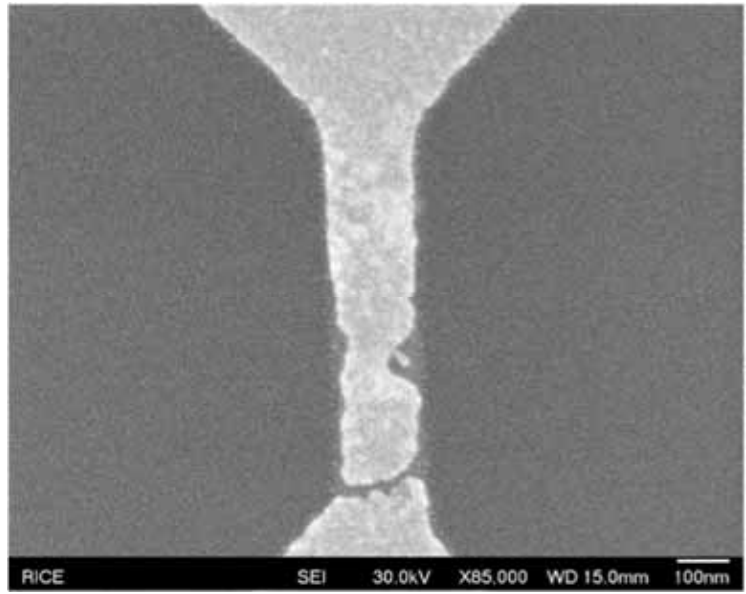

**Figure S4**. SEM image of the device demonstrated in the Figure 2 of the main text after the formation of the nanogap by electromigration at room temperature.

## 3. Additional experimental data for devices with nanogaps formed by electromigration

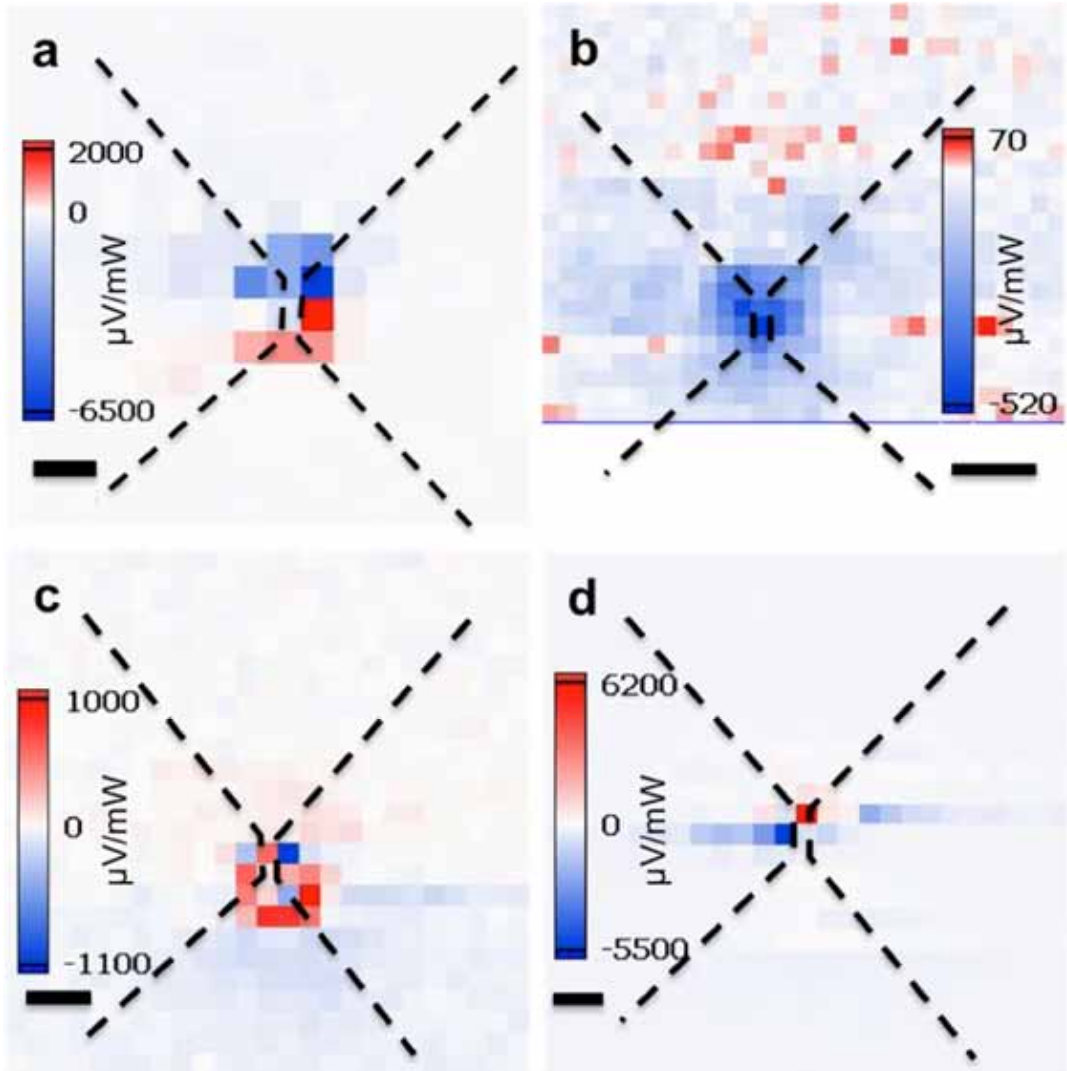

**Figure S5.** Additional examples of the enhancement of the photovoltage map after the formation of the nanogap. Au devices with Ti adhesion layer: (a) Device resistance is 0.8 MOhm, substrate temperature is 5.2 K. Laser power on the sample is 8 mW; (b) Device resistance is 110 kOhm, substrate temperature is 5.2 K. Au devices without Ti adhesion layer: (c) Device resistance is 0.9 MOhm, substrate temperature is 5.2 K. (d) Device resistance is 0.8 MOhm, substrate temperature is 293 K. Units in all panels are in µV per mW of laser power on the sample. Scale bars are 1 µm.

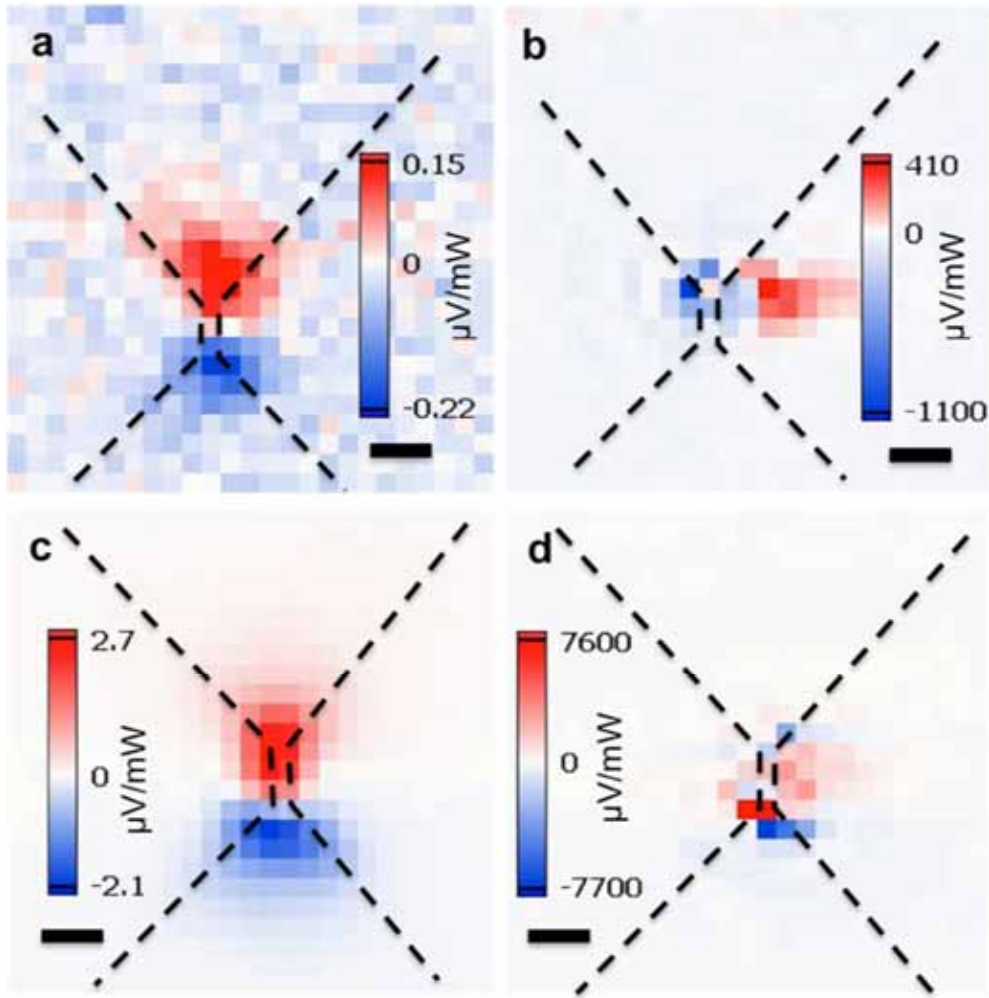

**Figure S6.** Additional examples of the enhancement of the photovoltage map after the formation of the nanogap during the Au device electromigration at room temperature. (a) Device fabricated with Ti adhesion layer, initial PTE voltage map. (b) After the nanogap formation. Total device resistance is 0.13 GOhm. (c) Device fabricated without Ti adhesion layer, initial PTE voltage map. (d) After the nanogap formation, resistance is 13 GOhm. Units in all panels are in μV per mW of laser power on the sample. Scale bars are 1 μm.

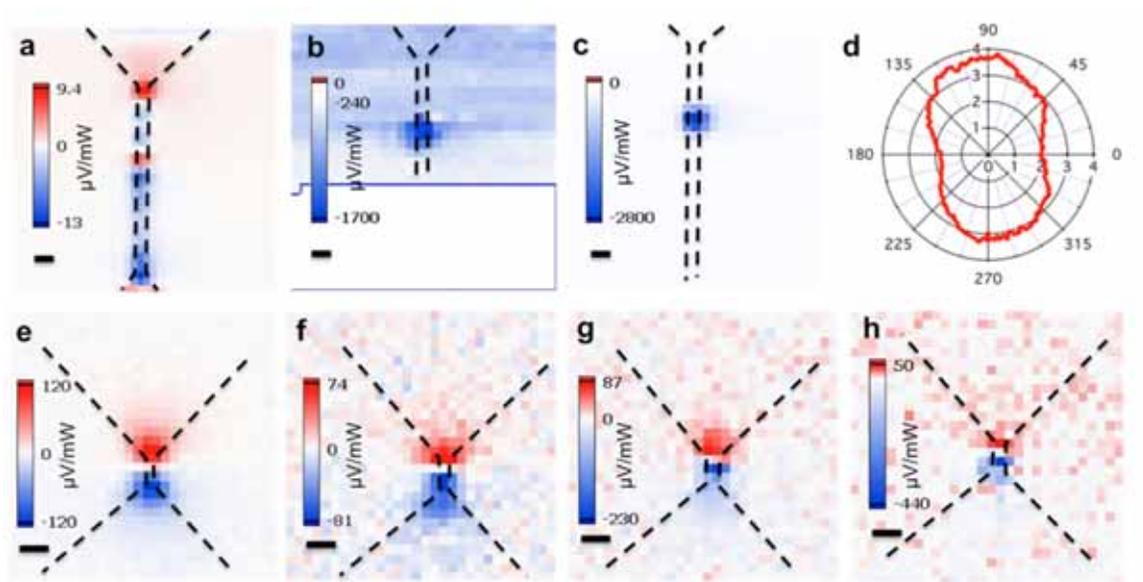

**Figure S7.** Evolution of the PTE voltage maps for AuPd devices fabricated with (top row) and without (bottom row) Ti adhesion layer as the devices are electromigrated to form a nanogap. (a) Initial distribution of PTE voltage for an unbroken device measured at room temperature. (b) The same after the formation of the nanogap with 20 MOhm resistance. (c) The same device after cooling down to the base temperature of 5 K. Resistance of the nanogap is increased to 0.6 GOhm. (d) Polarization dependence of the magnitude of the PTE voltage recorded at the minimum point from the previous panel. (e) Initial PTE voltage map for AuPd device without Ti adhesion layer. Initial resistance is 100 Ohm. The increased magnitude of the signal relative to Au devices is due to a larger Seebeck coefficient of AuPd alloy. The experiment is performed at base temperature of 5 K to enhance the stability of the device at intermediate resistance values. (f) Nanoscale constriction is formed with resistance of 0.7 kOhm. (g) Resistance is increased to 5.6 kOhm. (h) Photovoltage map after the nanogap formation with resistance of 36 kOhm. This device was further electromigrated to 36 MOhm and displayed enhanced photovoltage with ~1000 µV per mW of laser power on the sample (these units are used for all panels except for (e) in which the units are changed to mV per mW of laser power).

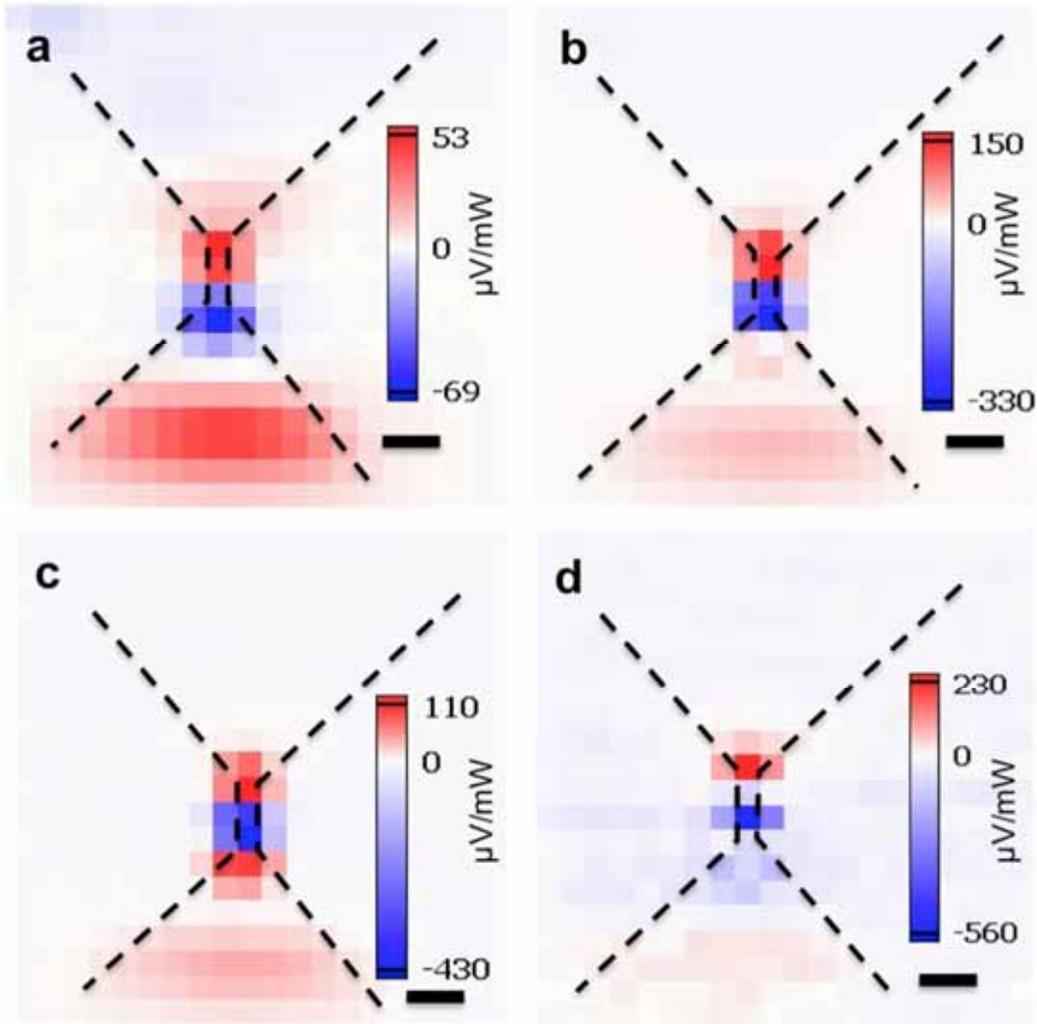

**Figure S8.** Evolution of the photovoltage maps for Ni devices fabricated without a Ti adhesion layer as the devices are electromigrated to form a nanogap at substrate temperature of 5 K. Device resistance is progressively increased: 190 Ohm (a), 744 Ohm (b), 2.9 kOhm (c), 48 kOhm (d). Units in all panels are in µV per mW of laser power on the sample.

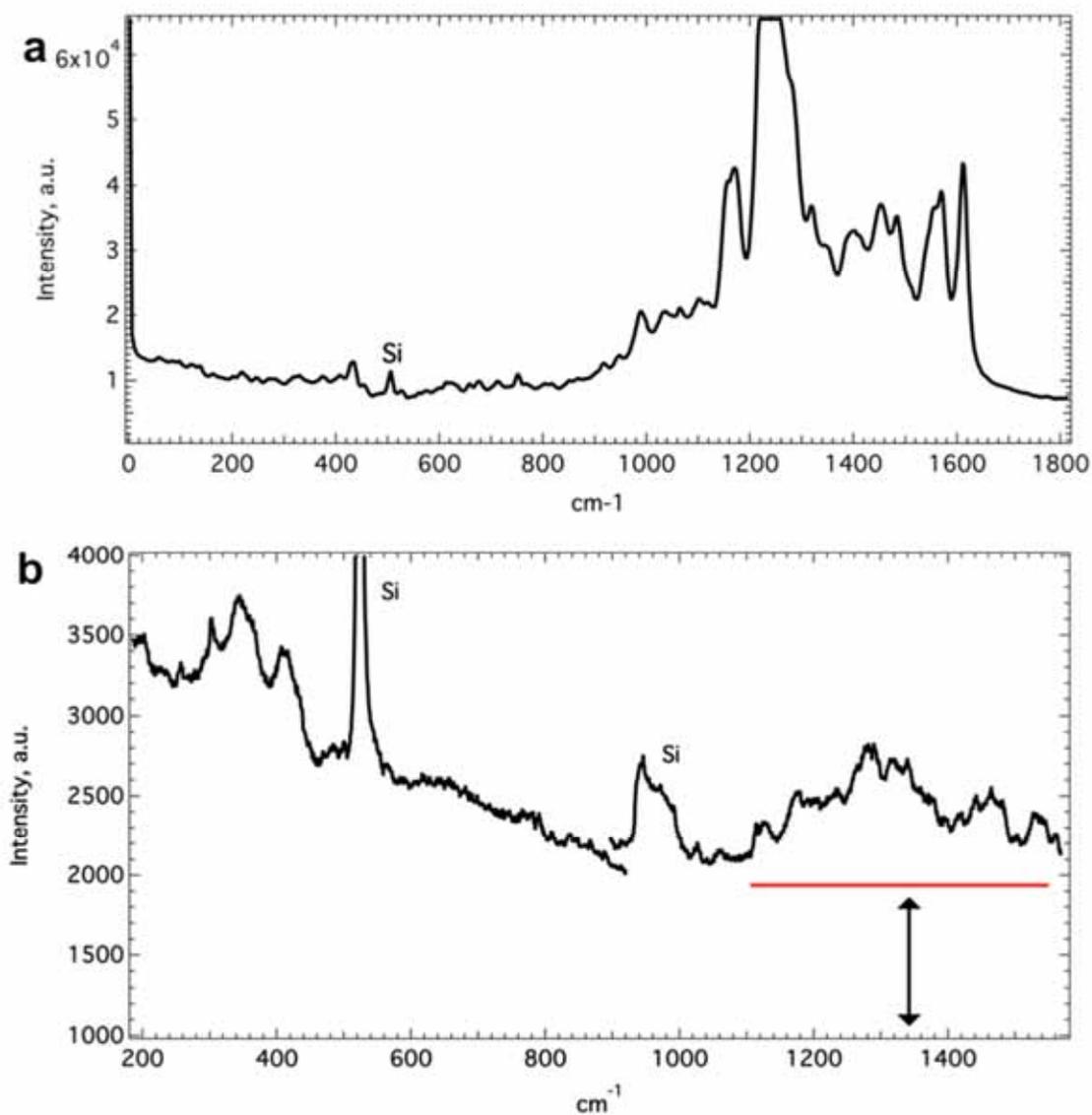

**Figure S9.** Raman spectra of the adsorbed contaminants acquired after the nanogap formation. (a) Spectrum acquired during the photovoltage map measurement for the device demonstrated in Figure 3e of the main text. (b) The same as in (a), but for Figure S5a. Red line demonstrates the Raman background increase after the nanogap formation. Lines that correspond to silicon substrate are labeled as Si.

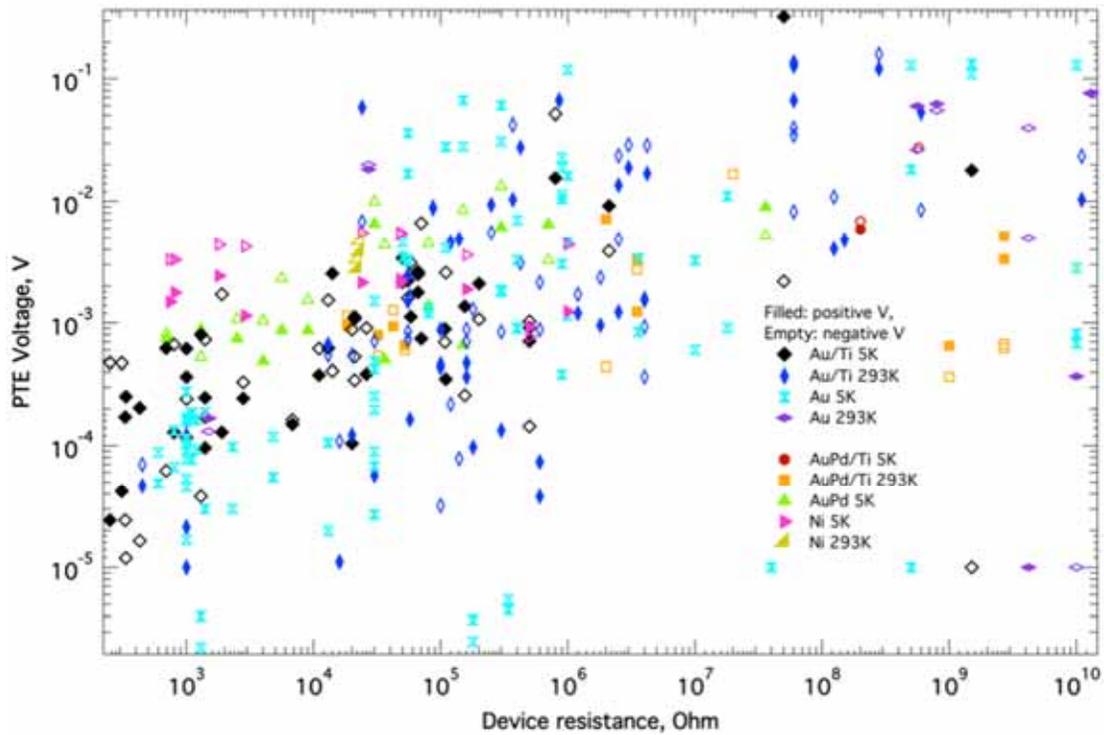

**Figure S10.** Compilation of the photovoltages for different devices acquired at laser intensity of 10mW as a function of the device resistance after the nanoscale constriction is formed by electromigration. Legend: filled (empty) markers correspond to the maximums (minimums) in the photovoltage map for a particular device; Au/Ti (Au) denotes gold devices with (without) Ti adhesion layer; AuPd/Ti (AuPd) denotes devices with (without) Ti adhesion layer; Ni denotes devices fabricated from Ni without Ti adhesion layer; 5K and 293K denotes the substrate temperature at which the photovoltage maps were acquired. Data displays the overall trend for the larger photovoltages as the device resistance is increased with the slope of ~1nA until the resistance reaches $R\sim10^7\Omega$, after which the average photovoltage is ~10mV.

# 4. Discussion of the possible mechanisms that can contribute to the generation of the photovoltages in plasmonic nanogaps

There are several possible physical mechanisms that can contribute to generation of the enhanced photovoltages in nanowires with atomic-scale constrictions and plasmonic nanogaps. Conventional Seebeck physics of the electrode material seems to be ruled out by the magnitude of the observed photovoltage. In experiments on atomic scale junctions and point contacts with bulk heaters, the thermoelectric power of the nanoscale constriction was demonstrated to be reduced relative to the bulk metal and dominated by the energy dependent transmittance of the atomic-scale contact.[3,20,21] The Seebeck coefficient for tunneling nanogaps remains small ~10 µV/K and weakly dependent to the junction parameters.[13] This suggests that the enhanced photovoltage as the device is electromigrated to form a constriction, but prior to nanogap formation, is due to the change in the local temperature as the thermal transport between two sides of the nanostructure is reduced. After the nanogap formation a number of additional photophysical processes can contribute to the photovoltage.

Optical rectification in plasmonically active nanogaps can generate dc photocurrents in devices shorted by low resistances.[22–24] Rectification requires non-linearity of the underlying *I-V* curve, and under illumination can produce a zero-bias short-circuit current proportional to $d^2I/dV^2$ at $V = 0$. Under nominally open-circuit conditions, this would be expected to produce a photovoltage proportional to the rectification current and nanogap resistance. The capacitance of the electrodes and the relative rate of dc charge pumping and charge relaxation will determine the steady state photovoltage. The sign of the open-circuit photovoltage produced by charge pumping from optical rectification should not depend on the position of the beam spot, unlike what we observe, and there is no correlation between observed photovoltage and $d^2I/dV^2$ measured in these junctions. We therefore do not believe the optical rectification to be the dominant mechanism that produces enhanced PTE voltages.

A transient electrical current across the nanogap created by the plasmon enhanced photon-assisted tunneling can produce transient voltages.[25] This transient voltage across the gap could create an artifact signal in the lock-in synchronized at the same frequency as the chopping of the incident radiation. The magnitude of the photovoltage generated by this mechanism is smaller than expected from tunneling of high-energy carriers.[26] We also dismiss a possible contribution from the recently demonstrated plasmoelectric[27] mechanism for photovoltage production in plasmonic nanostructures, because such a response should not require interelectrode tunneling conduction, and we do not observe such a photovoltage in nanostructures that were specially designed to enhance this effect.

Another possible source of enhanced PTE response could be the Seebeck coefficient of adsorbates or contaminants in the nanogap. As seen in previous studies, after nanogap formation the transverse dipolar mode in the nanowire strongly couples to "dark" multipolar plasmon modes that emerge from the asymmetries in the local geometry of the nanogap, resulting in a local field enhancement sufficient for SERS.[28] In addition to the PTE voltage measurement, we record Raman spectra at each point of the scan to locate the position of the nanowire and to assess the presence of the adsorbate molecules in the nanogap after electromigration. All devices displayed measurable Raman spectra in the 1100-1600 cm$^{-1}$ spectral region after the nanogap formation. The intensity and specific peaks positions vary between devices and are not correlated to the magnitude of the enhanced PTE voltage, Figure S9. These molecular adsorbates and TiO$_x$ that can be potentially present in the nanogap for the devices with Ti adhesion layer could contribute to the residual conductance after nanogap formation, and to the observed PTE voltage. In a composite nanostructure of graphene overlaid with the plasmonic nanogap bowtie antenna, no photovoltage was observed if the graphene layer was absent.[24] In our experiments the enhanced photovoltage is typically accompanied by SERS, which we attribute to the adsorbed organic molecules and is most consistent with Raman spectra of hydrogenated amorphous carbon.[29] Using an exaggerated (and therefore conservative) estimate of Seebeck coefficient of ~100 µV/K for this unknown organic material and a temperature increase of ~ 10 K across the nanogap for a room temperature experiment we arrive at maximum contribution of ~1-2mV, which is at least a factor of 10 lower than the photovoltage routinely observed in devices with plasmonic nanogaps.[30,31] Moreover, contaminant Seebeck response alone would not explain the extreme spatial localization of the large photovoltages. We therefore conclude that direct Seebeck contribution from organic contamination is unlikely to be the major mechanism at work. As we observe enhanced photovoltages in nanostructures fabricated without Ti adhesion layer, we also exclude the possible contribution from thermoelectric effects in non-stoichiometric $TiO_x$ despite the reports of very large, composition-dependent Seebeck coefficient.[32,33]

To elucidate the role of plasmons in the large photovoltage response, we fabricated devices of the same geometry, but from films of AuPd alloy and Ni. Taking into account the larger values of the Seebeck coefficient in AuPd and Ni, the unbroken devices display similar behavior as the Au devices, Figure S7, S8. After the formation of the nanogap, the photovoltage is also enhanced, however to a lesser degree than in the plasmonic Au devices. For the Ni devices the maximum photovoltage enhancement from unmigrated constriction to nanogap is ~10×. AuPd devices display a larger variability in enhancement factors ranging from ~4× for devices without Ti adhesion layer, to ~300× for devices with Ti adhesion layer. These control experiments suggest that the effect from the plasmon excitation in Au nanogap devices affects the photovoltage beyond simple enhanced steady-state heating.

## References

(1) Mahan, G. D. The Benedicks Effect: Nonlocal Electron Transport in Metals. *Phys. Rev. B* **1991**, *43*, 3945–3951.
(2) Bärner, K.; Morsakov, W.; Irrgang, K. Thermovoltages under a Non‐linear Seebeck Coefficient. *Phys. Status Solidi A* **2016**, *213*, 1553–1558.
(3) Ludoph, B.; Ruitenbeek, J. M. van. Thermopower of Atomic-Size Metallic Contacts. *Phys. Rev. B* **1999**, *59*, 12290–12293.
(4) Evangeli, C.; Matt, M.; Rincón-García, L.; Pauly, F.; Nielaba, P.; Rubio-Bollinger, G.; Cuevas, J. C.; Agraït, N. Quantum Thermopower of Metallic Atomic-Size Contacts at Room Temperature. *Nano Lett.* **2015**, *15*, 1006–1011.

(5) Szakmany, G. P.; Orlov, A. O.; Bernstein, G. H.; Porod, W. Single-Metal Nanoscale Thermocouples. *IEEE Trans. Nanotechnol.* **2014**, *13*, 1234–1239.
(6) Szakmany, G. P.; Orlov, A. O.; Bernstein, G. H.; Porod, W. Novel Nanoscale Single-Metal Polarization-Sensitive Infrared Detectors. *IEEE Trans. Nanotechnol.* **2015**, *14*, 379–383.
(7) Szakmany, G. P.; Orlov, A. O.; Bernstein, G. H.; Porod, W. Shape Engineering of Antenna-Coupled Single-Metal Nanothermocouples. *Infrared Phys. Technol.* **2015**, *72*, 101–105.
(8) Sun, W.; Liu, H.; Gong, W.; Peng, L.-M.; Xu, S.-Y. Unexpected Size Effect in the Thermopower of Thin-Film Stripes. *J. Appl. Phys.* **2011**, *110*, 83709.
(9) Liu, H.; Sun, W.; Xu, S. An Extremely Simple Thermocouple Made of a Single Layer of Metal. *Adv. Mater.* **2012**, *24*, 3275–3279.
(10) Fuchs, K. The Conductivity of Thin Metallic Films according to the Electron Theory of Metals. *Math. Proc. Camb. Philos. Soc.* **1938**, *34*, 100–108.
(11) Sondheimer, E. H. The Mean Free Path of Electrons in Metals. *Adv. Phys.* **2001**, *50*, 499–537.
(12) Pichard, C. R.; Tellier, C. R.; Tosser, A. J. Thermoelectric Power of Thin Polycrystalline Metal Films in an Effective Mean Free Path Model. *J. Phys. F Met. Phys.* **1980**, *10*, 2009.
(13) Russer, J. A.; Jirauschek, C.; Szakmany, G. P.; Schmidt, M.; Orlov, A. O.; Bernstein, G. H.; Porod, W.; Lugli, P.; Russer, P. A Nanostructured Long-Wave Infrared Range Thermocouple Detector. *IEEE Trans. Terahertz Sci. Technol.* **2015**, *5*, 335–343.
(14) Das, V. D.; Soundararajan, N. Size and Temperature Effects on the Seebeck Coefficient of Thin Bismuth Films. *Phys. Rev. B* **1987**, *35*, 5990–5996.
(15) Lin, S. F.; Leonard, W. F. Thermoelectric Power of Thin Gold Films. *J. Appl. Phys.* **1971**, *42*, 3634–3639.
(16) Cattani, M.; Salvadori, M. C.; Vaz, A. R.; Teixeira, F. S.; Brown, I. G. Thermoelectric Power in Very Thin Film Thermocouples: Quantum Size Effects. *J. Appl. Phys.* **2006**, *100*, 114905.
(17) Crisp, R. S.; Rungis, J. Thermoelectric Power and Thermal Conductivity in the Silver-Gold Alloy System from 3-300°K. *Philos. Mag.* **1970**, *22*, 217–236.
(18) Ho, C. Y.; Bogaard, R. H.; Chi, T. C.; Havill, T. N.; James, H. M. Thermoelectric Power of Selected Metals and Binary Alloy Systems. *Thermochim. Acta* **1993**, *218*, 29–56.
(19) Zolotavin, P.; Alabastri, A.; Nordlander, P.; Natelson, D. Plasmonic Heating in Au Nanowires at Low Temperatures: The Role of Thermal Boundary Resistance. *ACS Nano* **2016**, *10*, 6972–6979.
(20) Shklyarevskii, O. I.; Jansen, A. G. M.; Hermsen, J. G. H.; Wyder, P. Thermoelectric Voltage Between Identical Metals in a Point-Contact Configuration. *Phys. Rev. Lett.* **1986**, *57*, 1374–1377.
(21) Weber, L.; Lehr, M.; Gmelin, E. Investigation of the Transport Properties of Gold Point Contacts. *Phys. B Condens. Matter* **1996**, *217*, 181–192.
(22) Ward, D. R.; Hüser, F.; Pauly, F.; Cuevas, J. C.; Natelson, D. Optical Rectification and Field Enhancement in a Plasmonic Nanogap. *Nat. Nanotechnol.* **2010**, *5*, 732–736.

(23) Arielly, R.; Ofarim, A.; Noy, G.; Selzer, Y. Accurate Determination of Plasmonic Fields in Molecular Junctions by Current Rectification at Optical Frequencies. *Nano Lett.* **2011**, *11*, 2968–2972.

(24) Shi, S.-F.; Xu, X.; Ralph, D. C.; McEuen, P. L. Plasmon Resonance in Individual Nanogap Electrodes Studied Using Graphene Nanoconstrictions as Photodetectors. *Nano Lett.* **2011**, *11*, 1814–1818.

(25) Ittah, N.; Noy, G.; Yutsis, I.; Selzer, Y. Measurement of Electronic Transport through 1G0 Gold Contacts under Laser Irradiation. *Nano Lett.* **2009**, *9*, 1615–1620.

(26) Fung, E.-D.; Adak, O.; Lovat, G.; Scarabelli, D.; Venkataraman, L. Too Hot for Photon-Assisted Transport: Hot-Electrons Dominate Conductance Enhancement in Illuminated Single-Molecule Junctions. *Nano Lett.* **2017**, *17*, 1255–1261.

(27) Sheldon, M. T.; Groep, J. van de; Brown, A. M.; Polman, A.; Atwater, H. A. Plasmoelectric Potentials in Metal Nanostructures. *Science* **2014**, *346*, 828–831.

(28) Herzog, J. B.; Knight, M. W.; Li, Y.; Evans, K. M.; Halas, N. J.; Natelson, D. Dark Plasmons in Hot Spot Generation and Polarization in Interelectrode Nanoscale Junctions. *Nano Lett.* **2013**, *13*, 1359–1364.

(29) Veres, M.; Füle, M.; Tóth, S.; Koós, M.; Pócsik, I. Surface Enhanced Raman Scattering (SERS) Investigation of Amorphous Carbon. *Diam. Relat. Mater.* **2004**, *13*, 1412–1415.

(30) Meyerson, B.; Smith, F. W. Thermopower of Doped Semiconducting Hydrogenated Amorphous Carbon Films. *Solid State Commun.* **1982**, *41*, 23–27.

(31) Takai, K.; Oga, M.; Sato, H.; Enoki, T.; Ohki, Y.; Taomoto, A.; Suenaga, K.; Iijima, S. Structure and Electronic Properties of a Nongraphitic Disordered Carbon System and Its Heat-Treatment Effects. *Phys. Rev. B* **2003**, *67*, 214202.

(32) Tsuyumoto, I.; Hosono, T.; Murata, M. Thermoelectric Power in Nonstoichiometric Orthorhombic Titanium Oxides. *J. Am. Ceram. Soc.* **2006**, *89*, 2301–2303.

(33) Tang, J.; Wang, W.; Zhao, G.-L.; Li, Q. Colossal Positive Seebeck Coefficient and Low Thermal Conductivity in Reduced TiO 2. *J. Phys. Condens. Matter* **2009**, *21*, 205703.